\documentclass[a4paper,10pt,oneside]{article}
\usepackage[latin1]{inputenc}
\usepackage{amsmath}
\usepackage{amsfonts}
\usepackage{amssymb}
\usepackage{graphicx}
\usepackage{footnote}
\usepackage{float}
\usepackage{subcaption}
\DeclareGraphicsExtensions{.bmp,.png,.pdf,.jpg,.eps}
\usepackage{physics}
\usepackage{mathrsfs}
\usepackage{cite}
\usepackage[colorlinks,urlcolor=blue,citecolor=blue]{hyperref}

\advance \textheight by \topskip
\usepackage{amsmath}
\usepackage[english]{babel}
\makeatletter
 \@addtoreset{equation}{section}
\makeatother
\usepackage[latin1]{inputenc}
\usepackage{amsmath}
\numberwithin{equation}{section}

\advance \textheight by \topskip
\usepackage{amsmath}
\usepackage[english]{babel}
\DeclareMathAlphabet\mathbfcal{OMS}{cmsy}{b}{n}
\DeclareMathAlphabet{\boldmathe}{T1}{cmr}{bx}{it}

\def\be{\begin{equation}}
\def\ee{\end{equation}}

\def\be{\begin{equation}}
\def\ee{\end{equation}}

\def\be{\begin{equation}}
\def\ee{\end{equation}}

\def\Ti{\text{i}}

\usepackage{todonotes}

\begin{document}

\begin{center}
{\LARGE \bf
	Lorentzian quantum wells in graphene: the role of shape invariance in zero-energy states trapping\\ 
}
\vspace{6mm}
{\Large Francisco Correa$^a$, Luis Inzunza$^{a,b}$ and V\'i{}t Jakubsk\'y$^c$ 
}
\\[6mm]

\noindent ${}^a${\em Departamento de F\'isica\\ Universidad de Santiago de Chile,  Av. Victor Jara 3493, Santiago, Chile}\\[3mm]
\noindent ${}^b${\em 
Instituto de Ciencias F\'isicas y Matem\'aticas\\ 
Universidad Austral de Chile, Casilla 567, Valdivia, Chile}\\[3mm]
\noindent ${}^c${\em
Nuclear Physics Institute\\ Czech Academy of Science, 250 68 \v{R}e\v{z}, Czech Republic}
\vspace{12mm}
\end{center}

\begin{abstract}
Confining Dirac fermions in graphene by electrostatic fields is a challenging task. Electric quantum dots created by a scanning tunneling microscope (STM) tip can trap zero-energy quasi-particles. The Lorentzian quantum well provides a faithful, exactly solvable, approximation to such a potential, hosting zero-energy bound states for certain values of the coupling constant. We show that in this critical configuration, the system can be related to the free particle model by means of a supersymmetric transformation. The revealed shape invariance of the model  greatly simplifies the calculation of the zero modes and naturally explains the degeneracy of the zero energy.  

\end{abstract}

\section{Introduction}
It is difficult to control Dirac fermions in graphene by electrostatic fields because the Klein tunneling allows the quasi-particles to escape from the electric traps. An electric field localized in one direction can form waveguides for Dirac quasi-particles with non-vanishing \cite{Tudorovski} or vanishing energy \cite{Downing3,Hartmann,Ho2}. Such localized electric dots can trap quasi-particles of non-vanishing energy only for finite time  \cite{Matulis}, but they can host bound states of zero energy \cite{Tudorovski,Downing,Ioffe0}. 

An axially symmetric electric field can be generated by an scanning tunnelling microscope tip positioned close to the graphene sheet. This configuration can be very well approximated by the Lorentzian well  potential \cite{Szafran,Mrenca,Downing, Downing2}. It allows the existence of analytical solutions of the radial Dirac equation \cite{Downing} in terms of hypergeometric functions. Moreover, the Lorentzian well potential can host square integrable eigenstates of zero energy, the zero modes, provided that its strength takes certain discrete values. For these values of the coupling constant, the hypergeometric functions are truncated into polynomials that render the wave functions square integrable. As a result, the zero energy degeneracy is always an even number and depends on the magnitude of the critical coupling constant. In the current article, we provide an alternative insight into the existence of zero modes in the Lorentzian well based on the supersymmetric transformations \cite{Cooper,Junker}. 

The supersymmetric (also called Darboux) transformations, known in the context of soliton theory \cite{Matveev}, can be used to construct new solvable quantum Hamiltonians from known ones. With this technique, it is possible to design and control the spectral properties of the new system, including the bound and scattering states. Thus, this method can be used to construct any physical quantity that we can obtain from the spectral/scattering data and wave functions. The supersymmetric transformations for one-dimensional Dirac-type operators \cite{Samsonov,Samsonov2} has been applied e.g. in the context of deformed nanotubes \cite{Jakubsky1}, integrable models \cite{Correa} or optical settings \cite{Correa2}. The zero modes in graphene have already been analyzed using supersymmetric methods in \cite{Ho,Schulze1,Ghosh}. In particular, these transformations were employed to reveal omnidirectional Klein tunneling of Dirac fermions of specific energy on various setups of analytical electrostatic scatterers \cite{cacj}. The supersymmetric formalism have been also successfully tested in condensed matter systems, for instance to obtain conductivity of the quantum Hall effect in graphene \cite{susyhall}. Additionally,  in \cite{waveguide1} it is suggested that the technique could be usefully in computing the electric conductance in waveguide problems \cite{waveguide2,waveguide3}.

The Darboux transformations are usually represented by a differential operator which provides a mapping between the eigenstates of the original and the new Hamiltonian. The similarities between the scattering properties of the original and the new systems are particularly evident in relativistic reflectionless systems. The reflectionless potential Hamiltonians are intertwined with the free particle Hamiltonian by such transformations. Since the free particle suffers no scattering, one can show that there is none in the new system. The potential term of this new system takes such a special form that backscattering is suppressed but, surprisingly, the bound states are still present. These quantum models also appear in the context of solitonic solutions in fermion field theories such as Gross-Neveu and Nambu-Jona-Lasinio \cite{Correa, Dunne:2013yra, dt2}. There is almost no limit to use this technique, any initial (solvable) Dirac or Schr\"odinger system can be modified in an infinite number of ways using the supersymmetric transformations. There is another interesting aspect of supersymmetric transformations worth mentioning here. The resulting potential of the transformed system may coincide with the potential of the original Hamiltonian up a different value of the couplings constants. If this is the case, the energy levels as well as the corresponding eigenstates can be found algebraically. Such systems are called shapeinvariant \cite{Dutt} and allows to classify solvable quantum systems. In the non-relativistic context, the P\"oschl-Teller, Scarf model as well as the harmonic oscillator are prominent examples in one dimension \cite{Cooper}.  In this article, we will show that these phenomena can also be found in two dimensional relativistic models of Dirac fermions confined by Lorentzian wells.

The article is organized as follows. In the next section, we review the exact solution of the stationary Dirac equation with Lorentzian well for zero energy. We clarify which are the critical values of the coupling parameter that support the existence of the bound states. In the section 3, we review the main features of the supersymmetric transformation for a generic Dirac operator described by $2\times 2$ matrices. In the section 4, we apply the Darboux transformation to the Lorentzian well potential and we show its shape invariant property and that the bound states can be generated by the transformation. 

\section{The Lorentzian well}

\begin{figure}
\begin{center}
\includegraphics[scale=0.13]{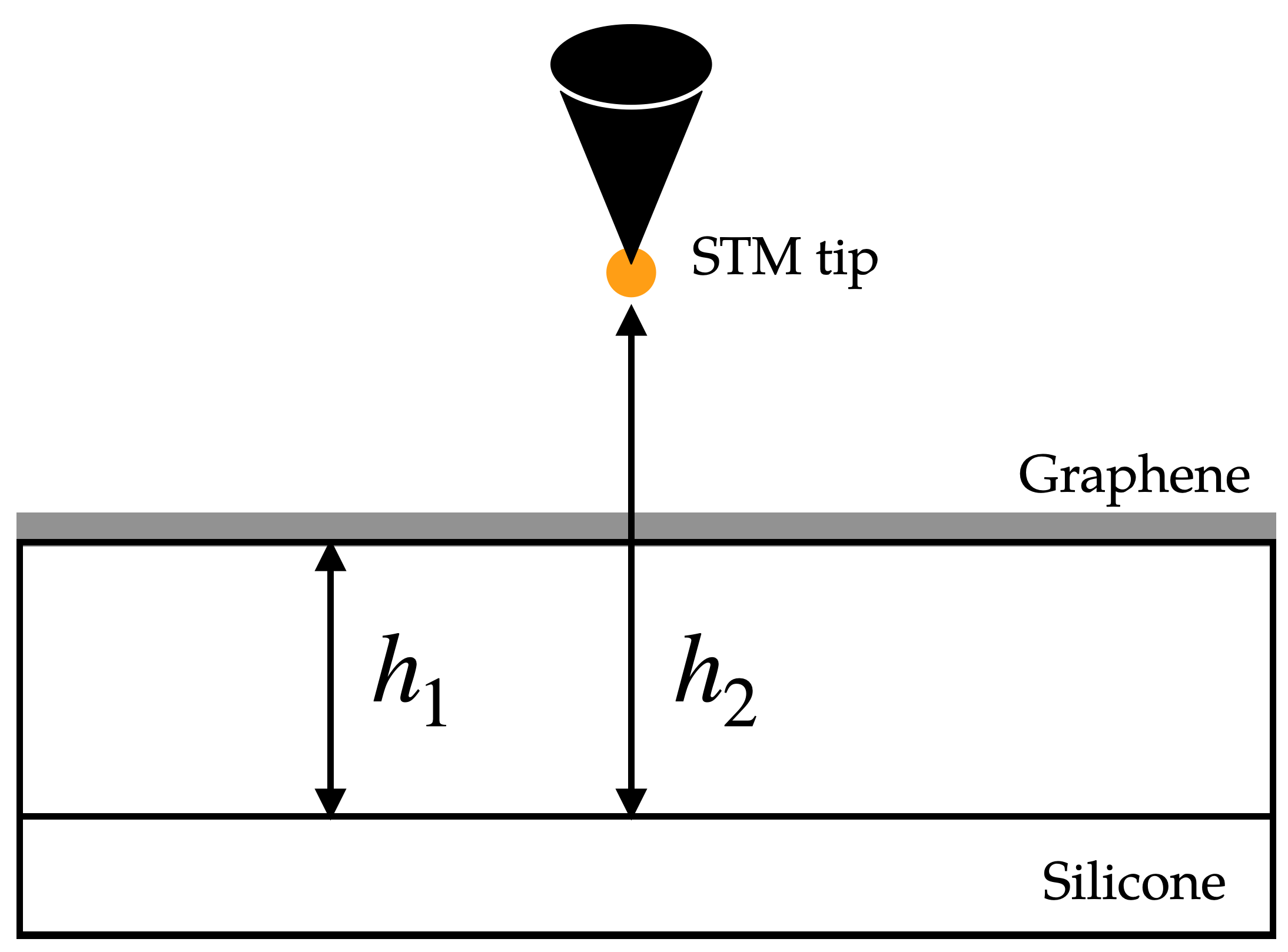} \qquad \qquad \qquad \qquad 
\includegraphics[scale=0.16]{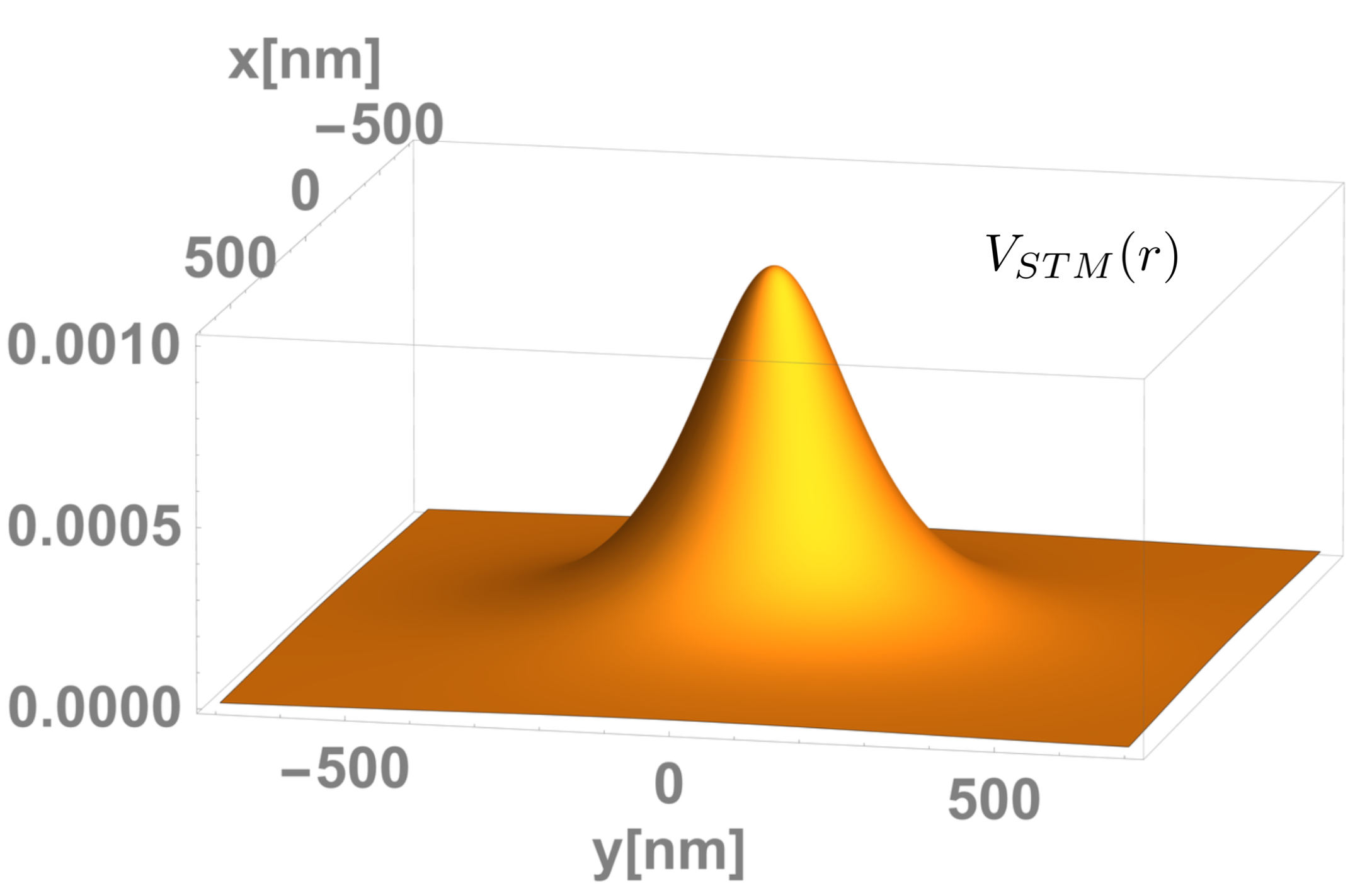}
\caption{a) Experimental setup of the Lorentzian well \cite{Downing}. The graphene monolayer is in gray and the silicon plate is in white.  b)  3-D Plot of the electric potential $V_{STM}(r)$ in the surface of the graphene layer with $h_1=20[nm]$, $h_2=200[nm]$ and $\frac{e Q}{4\pi \kappa}=1$.}
\label{Figtraj1}
\end{center}
\end{figure}

Let us suppose that the tip of scanning tunneling microscope (STM) is placed above a sheet of graphene. It is assumed to have rotational symmetry, which is rather good approximation of the real situation.  A silicon plate below the graphene is connected to the gate voltage and the STM tip, whose electric charge is denoted by $Q$. The distance between the silicon and graphene is $h_1$, whereas the distance between the STM tip and graphene is $h_2>h_1$, see Fig~\ref{Figtraj1} for illustration.   Thus, the setup can be understood by means of the standard charge image problem, which in the first order approximation leads to the following  effective electric potential at the graphene layer, 
\begin{equation}
\label{VSTM}
V_{STM}\approx\frac{e Q}{4\pi \kappa}\left(\frac{1}{\sqrt{r^2+(h_1-h_2)^2}}-\frac{1}{\sqrt{r^2+(h_1+h_2)^2}}\right)\, .
\end{equation}
Here, $\kappa$ is the Botzmann constant, and $r^2=x^2+y^2$ is the radial coordinate measured from the origin in the graphene layer, just below the tip, see Fig. \ref{Figtraj1}.   Let us take from the natural units where $\frac{e Q}{4\pi \kappa}=1$, although physical units will be recovered for discussion at the end of this section. 
The dynamics of an electron in such a configuration obeys the Dirac equation of a massless Weyl fermion coupled to the STM potential $V_{STM}$. Explicit analytical solutions of such equation are not known.
	
\begin{figure}[H]
\begin{center}
\includegraphics[scale=1.2]{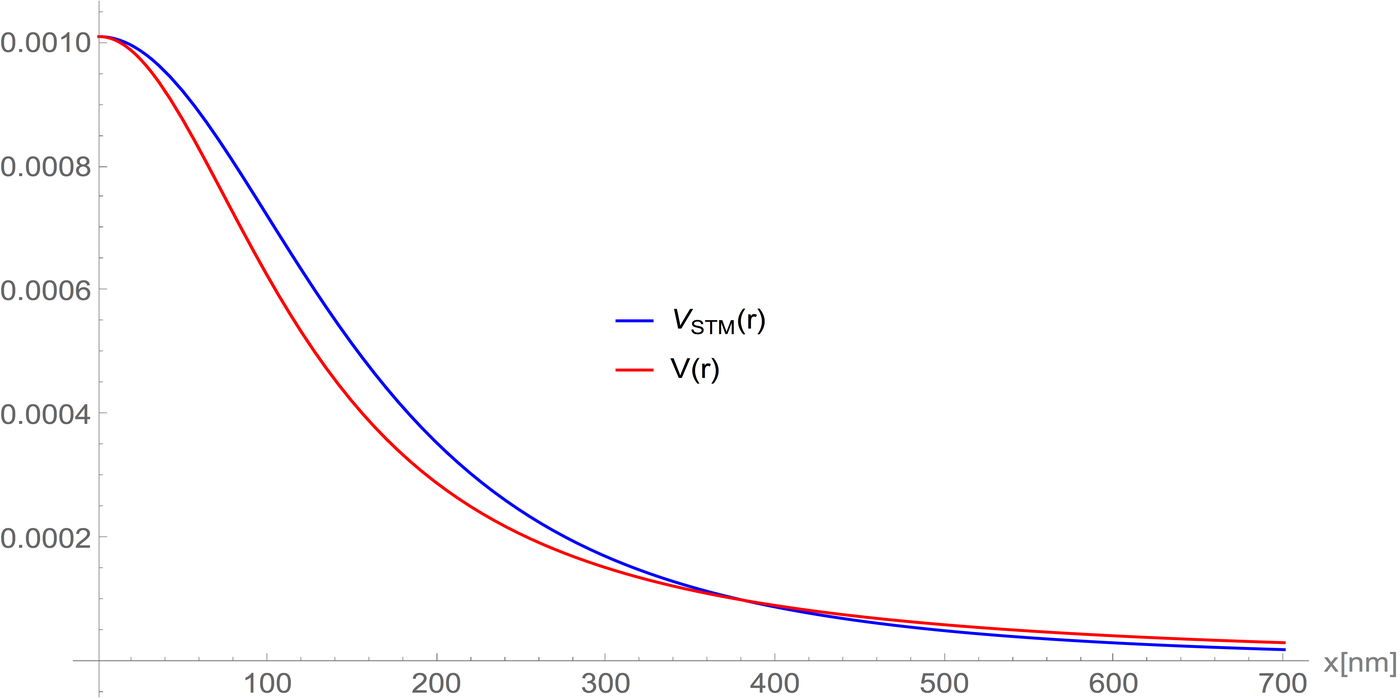}
\end{center}
\caption{\small{Comparative plots of the potentials $V_{STM}(r)$ (\ref{VSTM}) and $V(r)$ (\ref{Lor}). }}
\label{Fig3}
\end{figure}

It was argued in \cite{Downing} that the STM potential (\ref{VSTM}) has almost the same shape as the Lorentzian function 
\begin{equation}
\label{Lor}
V(r)=\frac{V_0}{1+(r/d)^2}\,,\qquad V_0= \frac{2h_1}{h_2^2-h_1^2}\,,\qquad
d=\frac{h_2^2-h_1^2}{\pi h_1}\ln(\frac{h_2+h_1}{h_2-h_1})\,.
\end{equation}
Therefore, it can be used as a faithful approximation of the realistic potential (\ref{VSTM}), see the comparative plots in Fig. \ref{Fig3}. The benefit of using (\ref{Lor}) instead of (\ref{VSTM}) stems from the fact that the Dirac equation with the ``Lorentzian well"   (\ref{Lor}) can be found analytically  \cite{Downing}.  This way, it was shown that there are bound states at zero energy for specific values of the coupling parameter $V_0$. Let us review these solutions due to their importance for the rest of the article.
Let us consider the zero energy equation of a two-dimensional massless fermion subject to the scalar Lorentzian well potential. The stationary Dirac equation reads as
\begin{equation}
\label{DiracLorentz}
H_\ell\psi=-\Ti(\sigma_1\partial_x+\sigma_2\partial_y)\psi+V_\ell(x,y)\sigma_0\psi=0\,,\quad V_\ell(x,y)= - \frac{2\ell}{1+x^2 +y^2},\quad \ell\in\mathbb{R},\quad \ell>0\,.
\end{equation}
In (\ref{DiracLorentz}), we consider only the coupling $\ell$ as positive since the Hamiltonian with $\ell<0$ can be obtained by a unitary transformation  $\sigma_3H_\ell\sigma_3=-H_{-\ell}$. The matrices $\sigma_i$, $i=1,2,3$, are the usual Pauli matrices and $\sigma_0$ is the identity matrix.
Notice that the Dirac Hamiltonian possesses rotational symmetry, despite the fact that the hexagonal lattice of graphene does not have the symmetry. The reason lies in the fact that free Dirac Hamiltonian provides approximation of tight-binding energy operator in regime of low energies where the dispersion relation  has almost linear dependence on momenta.
Therefore, the equation is separable in polar coordinates and can be solved explicitly  \cite{Downing}.

The regular solutions at the origin with positive angular momentum $m$ can be written as
\begin{align}
\label{spinorm}
\psi_{m,\ell}&=  \frac{e^{\Ti m \varphi } r^{m}}{(1+r^2)^{m+1}} \left(
\begin{array}{c}
{}_2F_1\left(m+1-\ell,m+1+\ell,1+m;\frac{r^2}{r^2+1}\right) \\
\Ti
\frac{ \ell r e^{\Ti \varphi }}{1+m} \,\,{}_2F_1\left(m+1-\ell,m+1+\ell,2+m;\frac{r^2}{r^2+1}\right) \\
\end{array}
\right)\,, \qquad m=0,1,2,\ldots\,,
\\
J\psi_{m,\ell} &=\left(m+\frac{1}{2}\right) \psi_{m,\ell}\,,\qquad
J=-\Ti \partial_\varphi+\frac{1}{2}\sigma_3\,.
\end{align}
The wavefunctions are given in terms of the hypergeometric function ${}_2F_1\left(a,b,c;z\right)$ with parameters $a,b$ and $c$ \cite{SpecialFunctions}. The case of solutions having negative values of the angular momentum $m$ can be obtained by the action of the symmetry operator $\mathcal{T}\sigma_2$, $[H_\ell,\mathcal{T}\sigma_2]=0$, where $\mathcal{T}$ is the complex conjugation operator, 
\begin{equation}
J(\mathcal{T}\sigma_2\psi_{m,\ell})=-\left(m+\frac{1}{2}\right)(\mathcal{T}\sigma_2\psi_{m,\ell})\,.
\end{equation} 

\begin{figure}[H]\qquad \quad
  \includegraphics[scale=0.8]{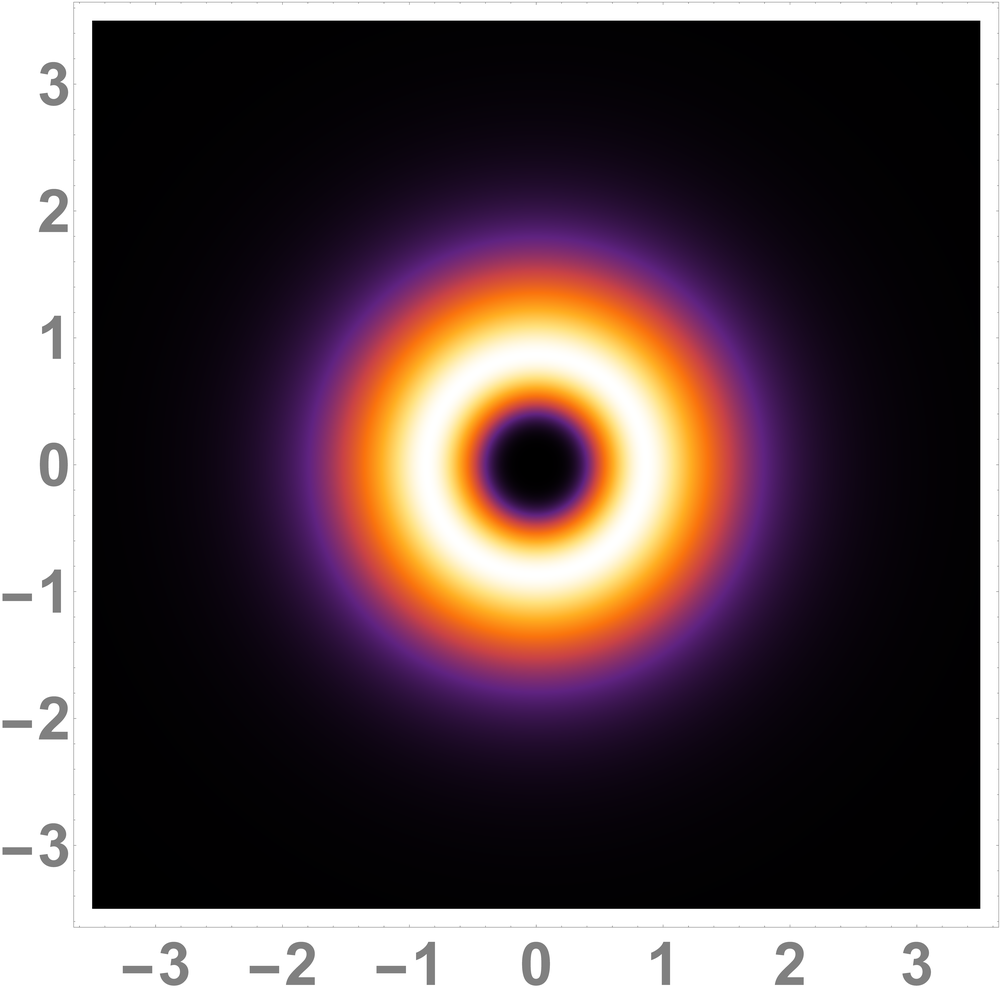}
 \includegraphics[scale=0.8]{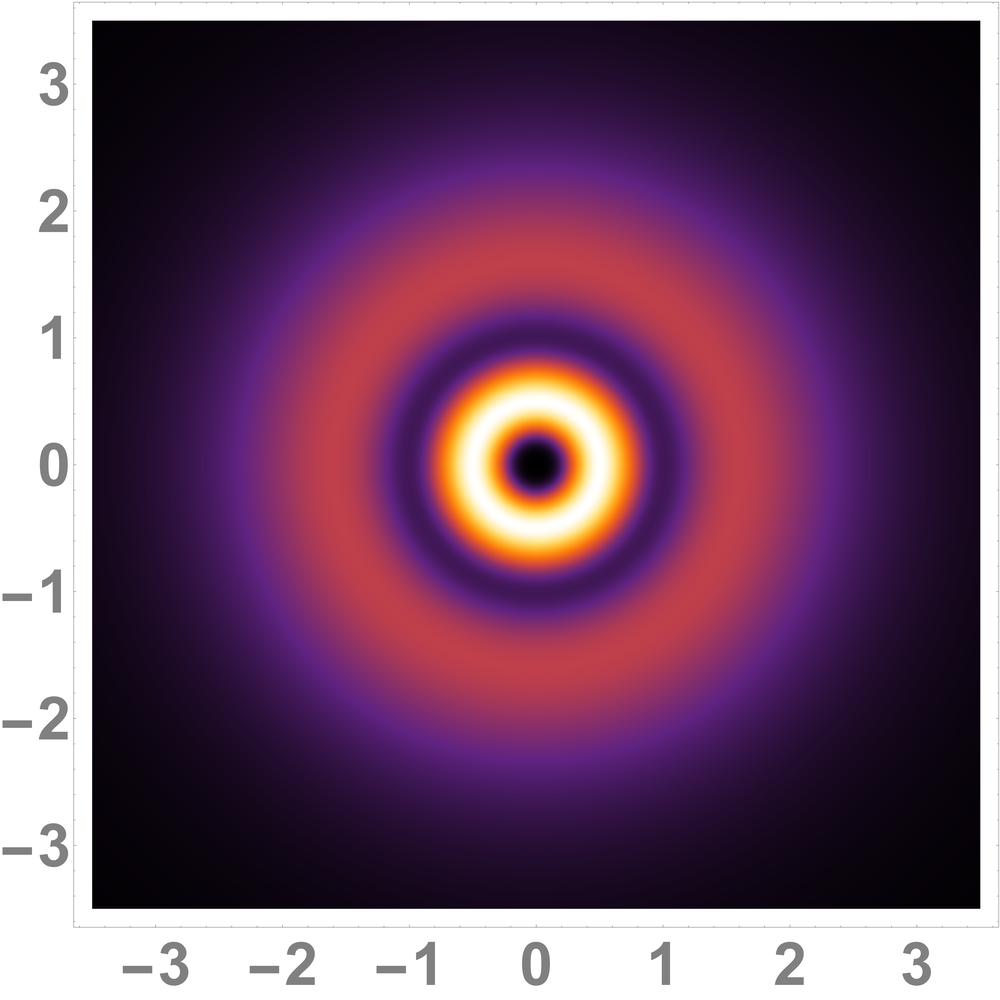}
 \includegraphics[scale=0.8]{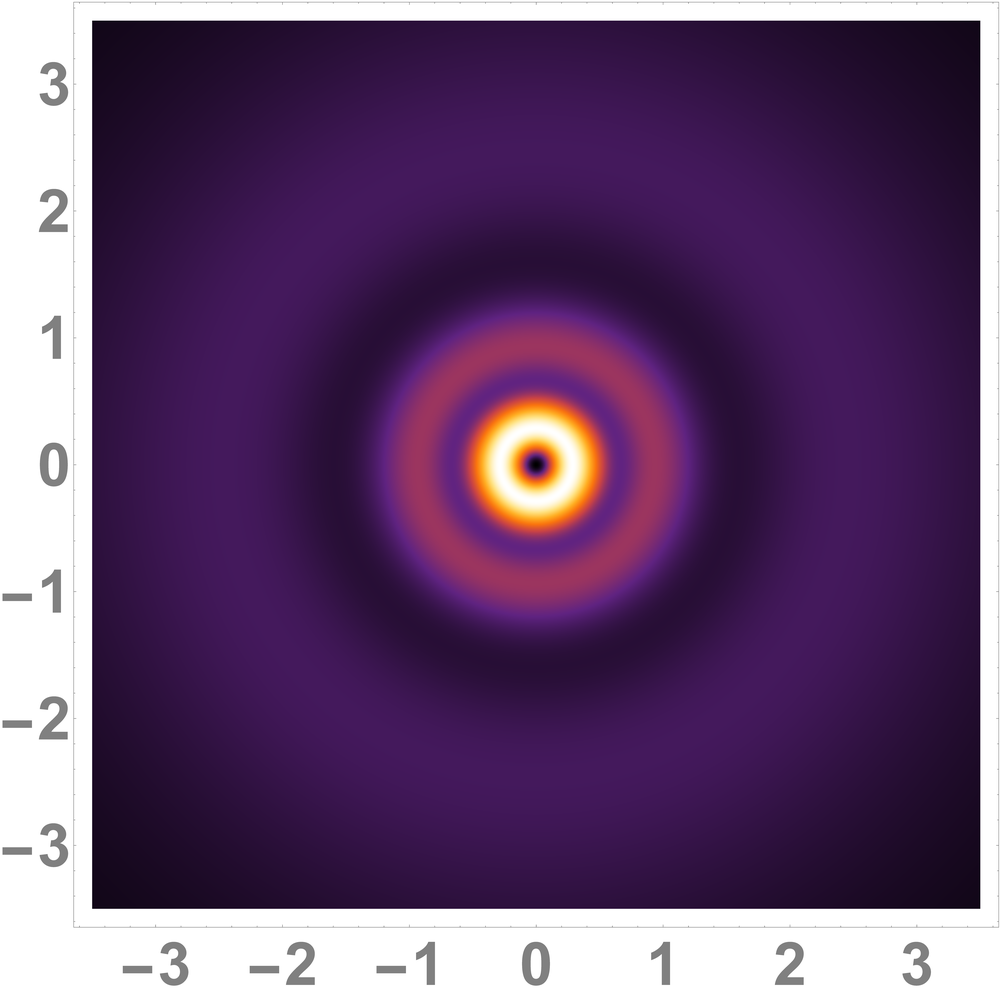}
 \includegraphics[scale=0.78]{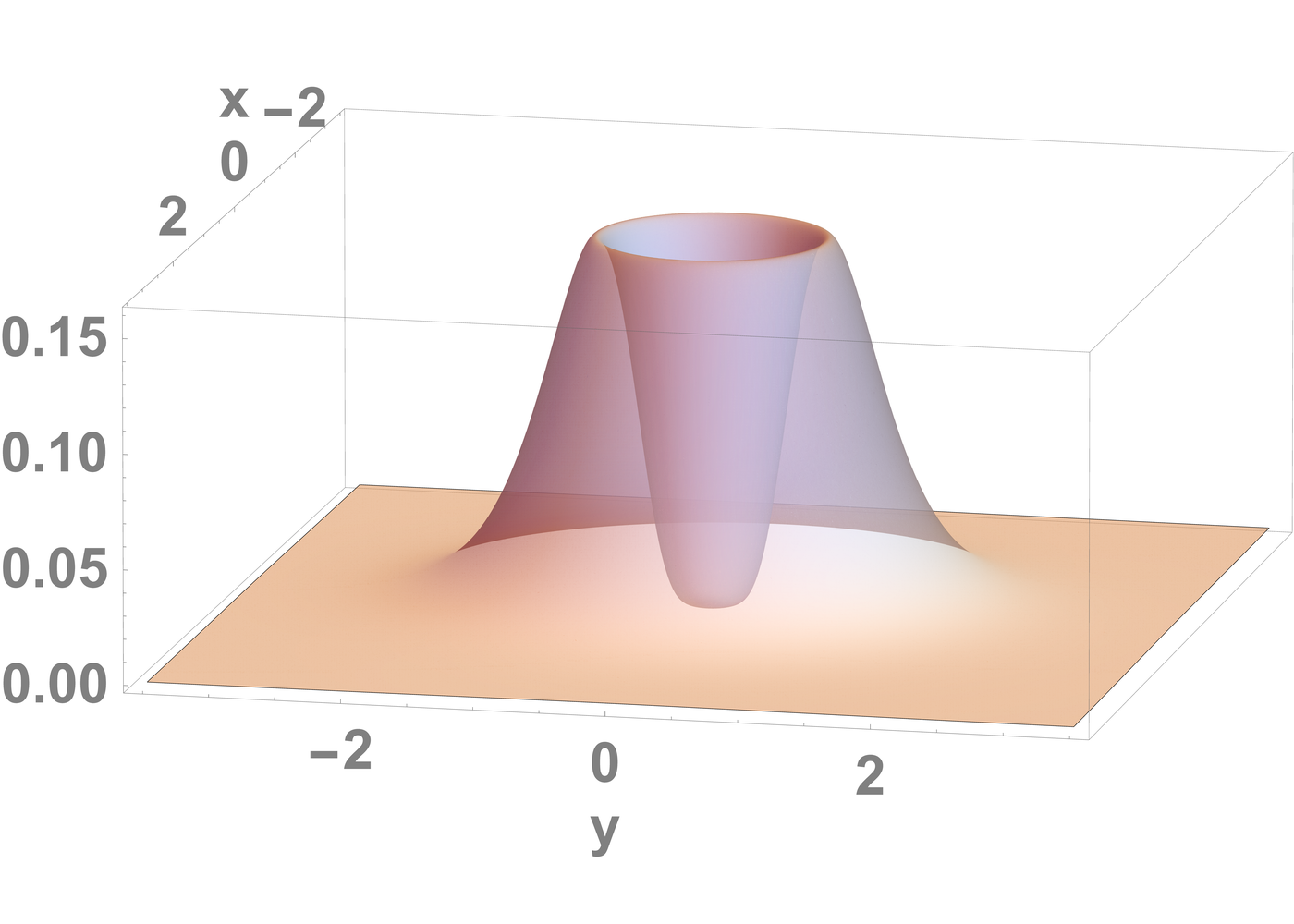} 
 \includegraphics[scale=0.78]{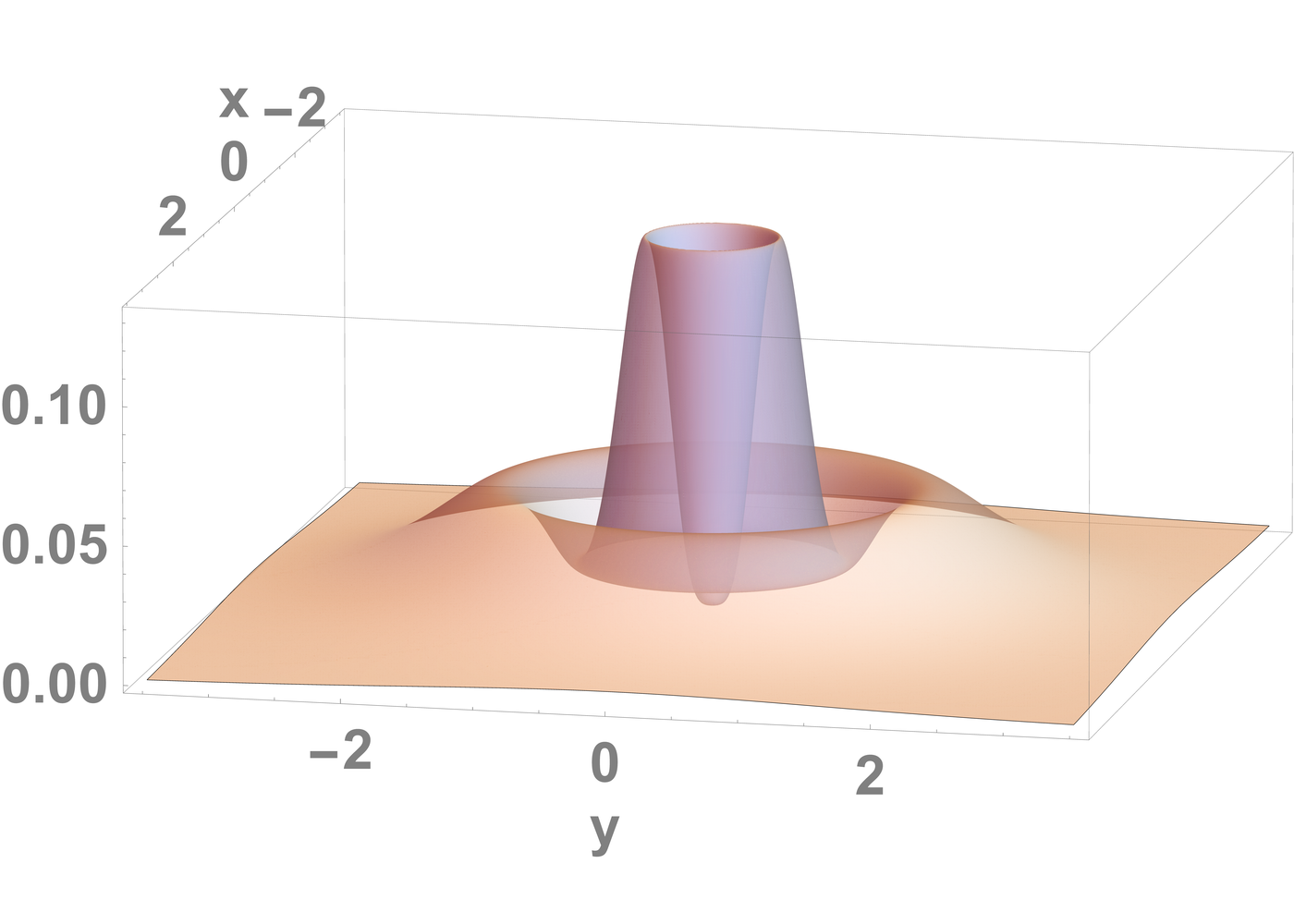}
 \includegraphics[scale=0.78]{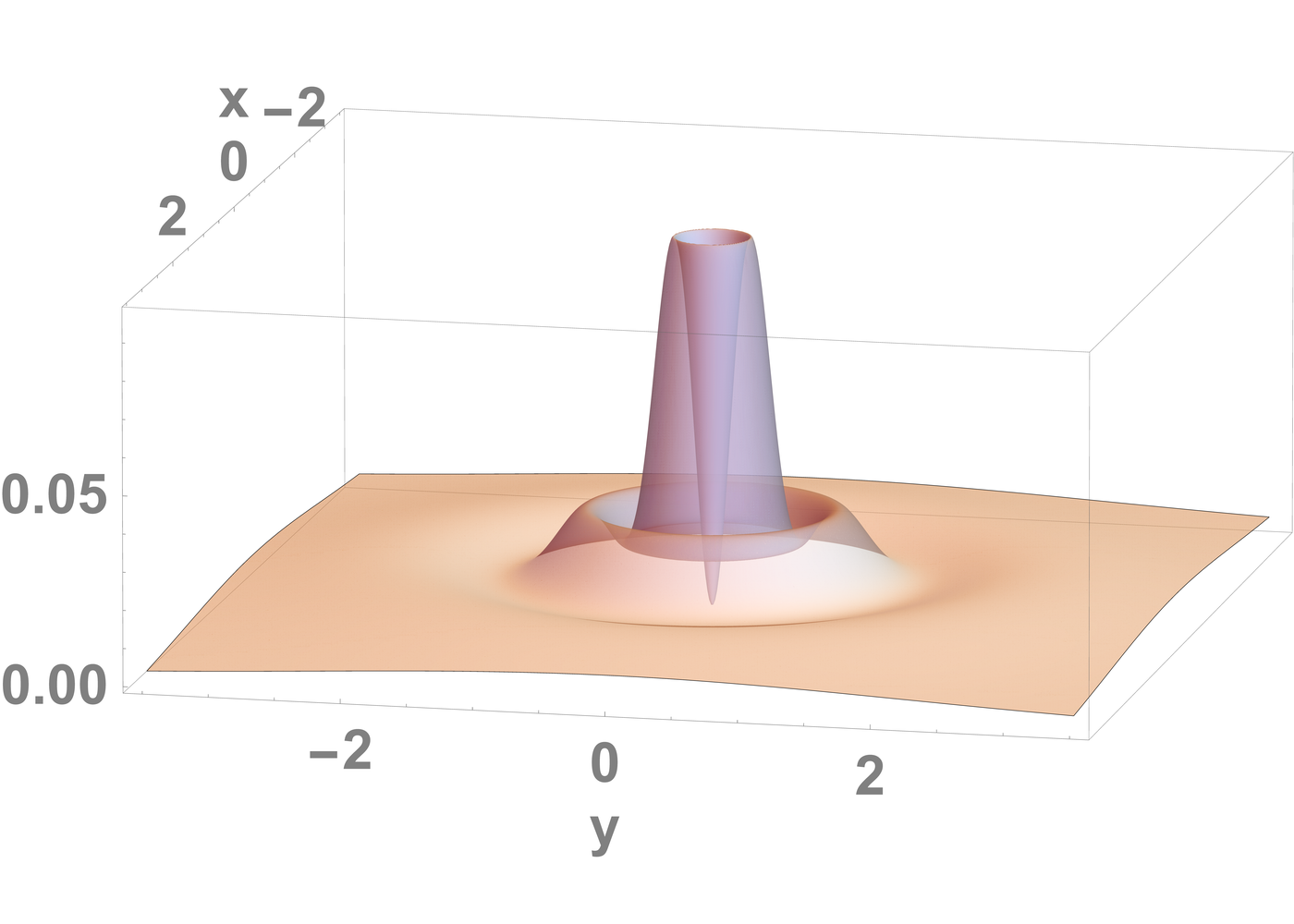}
 \caption{\small{
Plots of the probability density $\rho_{m,\ell}=\frac{\psi_{m,\ell}^\dagger\psi_{m,\ell}}{\bra{\psi_{m,\ell}}\ket{\psi_{m,\ell}}}$
for the case $\ell=4$. The normalizable state corresponds 
values $m=1,2,3$, form left to right (ordered respecting the number of nodes).
}}
\label{PlotsDen3}
 \end{figure}

The states (\ref{spinorm}) show a square integrable behavior provided that the hypergeometric functions are truncated to the Jacobi polynomials \cite{SpecialFunctions}. This is only possible if  either their first or the second argument is a negative integer. 
Remembering that $m$ is integer-valued (\ref{spinorm}), this implies that the coupling constant $\ell$ must also be an integer number.  In addition, square integrability implies that the angular momentum $m$ can only take  values from the restricted set $m={0,1,\ldots,\ell-1}$. Thus, the physical meaning of the components of $\psi_{m,\ell}$ can be assessed from their asymptotic behavior at the origin and at the infinity,
\begin{align}
\label{apsy}
r\rightarrow 0\,,&\quad
\psi_{m,\ell}\rightarrow^{} \left(\begin{array}{cc}
r^{m}\\
r^{1+m}
\end{array}\right)
\,,\qquad
&
r\rightarrow \infty\,,\quad  \psi_{m,\ell}\rightarrow 
\frac{1}{r^{1+m}}
\left(
\begin{array}{c}
\frac{1}{r}\\
1
\end{array}\right)
\,.
\end{align}
It is learned from here that $\psi_{m,\ell}$ are normalizable except for $m=0$, where logarithmic divergence emerges for large $r$. In conclusion, \textit{zero modes can only exist for integer values of $\ell$. The degeneracy of the zero energy states is an even number equal to $2(\ell-1)$ with the corresponding zero modes $\psi_{m,\ell}$ and $\mathcal{T} \sigma_2\psi_{m,\ell}$, where $m=1,\ldots\ell-1$.}
In Fig. \ref{PlotsDen3} we show how probability densities associated to normalized states 
 behave respect to $r$ in the case $\ell=4$. 
 
To recover the physical units in equation (\ref{DiracLorentz}), one should perform the scaling transformation  $(x,y)\rightarrow d(x,y)$, with $d$ in (\ref{Lor}), multiply the entire equation by $\hbar v_F$, where $v_F$ is the Fermi velocity and made the following identification,
\begin{equation}
\label{identification}
V_0=\frac{2\ell}{d}\hbar v_F= \frac{eQ}{4\pi \kappa} \frac{2h_1}{h_2^2-h_1^2}\,,\qquad
\ell=\frac{eQ}{4\pi^2 \kappa \hbar v_F} \ln(\frac{h_2+h_1}{h_2-h_1})\,.
\end{equation}  
The last equation in (\ref{identification}) means that for the realistic conditions in an experimental setup, the  parameters  $h_1$, $h_2$ and $Q$  must be fine-tunned to get an integer-valued $\ell$. In the same way, if the potential $V(r)=\frac{V_0}{1+(r/d)^2}$ has generic parameters $V_0$ and $d$, using a scale transformation of $r$ and the appropriate redefinition of $\ell$, we can obtain exactly the equation for the zero modes as in Eq. (\ref{DiracLorentz}). In this sense, different values of $d$ are translated in zoom in or zoom out the plots. The zero modes trapped by the Lorentzian well are bound states in the continuum, BICs. It has been argued in the literature \cite{Hsu} that BICs are fragile with respect to perturbations and turn into quasi-bound states (long-life resonances).  It seems to be difficult to detect BICs experimentally, since they do not couple with scattering states.   Nevertheless, some theoretical results \cite{Hsu,Gonzalez,Cortes} on graphene nanoribbons indicate that their detection might be possible with combined measurements of local density of states and conductance. This problem is also widely discussed in the experimental literature. There are already observations indicating that confinement is possible, at least at the level of quasi-bound states. Many of these experiments with satisfactory results take into account variations of the ideal setup in Fig. \ref{Figtraj1}, using doped graphene layers \cite{JLee,Matulis}, or  introducing external magnetic fields \cite{Freitag}. For relevant results on this topic see, \cite{LiLin,Yung} and references therein. 

The behavior of relativistic particles in the presence of Lorentzian wells is quite different from the non-relativistic one described by the Schr\"odinger Hamiltonian $H_{\text{Sch}}=-\partial_x^2-\partial_y^2-V_\ell(x,y)$.  In contrast to the relativistic case discussed here, to our knowledge there are no analytical results for the Schr\"odinger case. However, following the variational approach, one can conclude that Lorentzian wells can trap non-relativistic particles of negative energy for a wide range of coupling constants $\ell$, which are not restricted to integer values as in the relativistic case (see Appendix). Thus, it seems that Lorentzian wells are better at trapping non-relativistic particles than the relativistic ones. 

\section{Supersymmetric (Darboux) transformations in 2D Dirac systems}
\label{SecDar}
Darboux transformations make it possible to map a differential equation associated with a Hamiltonian system into another one while preserving the form of the kinetic term. In this way, the original Dirac (Schr\"odinger) Hamiltonian is transformed into a new Dirac (Schr\"odinger) one with a modified potential term. The solutions of the new system can be easily derived from the solutions of the original model. This type of transformation was originally studied in the context of differential equations by G. Darboux \cite{Darboux} 
before the advent of quantum theory.  Later on, the Darboux transformation found their application in supersymmetric quantum mechanics \cite{Cooper, Junker}. It is therefore often called a supersymmetric transformation. Here we briefly review the construction of Darboux transformations for a Dirac-type two-dimensional operator, see also \cite{Samsonov} and \cite{Samsonov2} for more details.

Let us suppose that we start with the original equation of the following form, 
\begin{equation}
\label{hgen} h\psi=(\gamma_1\partial_x+\gamma_2\partial_y+V(x,y))\psi=0\,,
\end{equation}
where both $\gamma_a$ and $V(x)$ are $2\times2$ matrices. The  $\gamma_a$ matrices have constant entries whereas the entries  of $V(x)$ can be arbitrary functions in general. At this point, the operator $h$ does not have to be hermitian. We assume that two solutions $\chi=\left({}^{\chi_1}_{\chi_2}\right)$ and $\xi=\left({}^{\xi_1}_{\xi_2}\right)$  of (\ref{hgen}) are known. Then, we construct the matrix $U$ whose columns correspond to $\chi$ and $\xi$,
\begin{equation}
h\chi=h\xi=0\, , \quad U=(\chi,\xi)=\begin{pmatrix}\chi_1&\xi_1\\
\chi_2&\xi_2\end{pmatrix}\,,\quad hU=0\,.
\end{equation}
Now, we define two additional operators $L$ and $\widetilde{h}$ in terms of the matrix $U$,
\begin{equation}\label{Ltildeh}
L=\partial_x-\Sigma\,,\qquad
\widetilde{h}=h- [\gamma_1,\Sigma]\,,\qquad
\Sigma=(\partial_xU) U^{-1}\,.
\end{equation}  
By construction, the ``seed" solutions $\chi$ and $\xi$ are annihilated by $L$, $L\chi=L\xi=0$. The operators $h$, $\widetilde{h}$ and $L$ satisfy the following formal equality
\begin{equation}\label{intgen}
L h=\widetilde{h}L\, .
\end{equation}
This is the so-called intertwining relation between the operators $h$ and $\widetilde{h}$, mediated by the differential operator $L$, which represents the Darboux transformation. It is an immediate consequence from (\ref{intgen}) that the solution of $h\psi=0$ can be transformed into the solution $\widetilde{\psi}$, $\widetilde{h}\widetilde{\psi}=0$,  
\begin{equation}\label{eq1}
h\psi=0\quad \Rightarrow\quad
\widetilde{h}\widetilde{\psi}=0,\quad \widetilde{\psi}=(L\psi) \, .
\end{equation}
It is worth noticing that both $L$ and $\widetilde{h}$ are uniquely defined in terms of the solutions $\chi$ and $\xi$. When compared to $h$, the operator $\widetilde{h}$ has an additional potential term $[\gamma,\Sigma]$ that is a function of $\chi$ and $\xi$, see (\ref{Ltildeh}). So far there are no specific requirements for the solutions and thus for the resulting Hamiltonian $\widetilde{h}$. However, it is clear that the explicit choice of the seeds $\chi$ and $\xi$ has a large impact on the form of $\widetilde{h}$. A common scenario is to construct the new systems devoid of singularities compared to $V(x,y)$, but this depends on the physics of the problem to be considered.

There can be constructed an ``inverse" intertwining operator $\widetilde{L}$, which mediates the inverse intertwining relation
\begin{equation}\label{intgen2}
\widetilde{L}\widetilde{h}=h\widetilde{L}\, .
\end{equation}
In the case that both $h$ and $\widetilde{h}$ are hermitian operators, we can write $\widetilde{L}=L^\dagger$. In the generic case, they can be formally written as $\widetilde{L}=\widetilde{U}\partial_x\widetilde{U}^{-1}$, where $\widetilde{U}$ is a specific eigenmatrix of $\widetilde{h}$,
\begin{equation}
\widetilde{h}\widetilde{U}=0 \, .
\end{equation} 
The columns $\widetilde{\chi}$ and $\widetilde{\xi}$ of the matrix $\widetilde{U}=(\widetilde{\chi},\widetilde{\xi})$ are sometimes called missing states in the literature. The name comes from the fact that, unlike $\widetilde{\psi}$ in (\ref{eq1}), they cannot be formally obtained by applying the $L$ operator to the eigenstates $\psi$ of the original Hamiltonian $h$. They are defined in terms the seed states $\chi$ and $\xi$ and may correspond to bound state of $\widetilde{h}$ that have no analog in the original equation (\ref{hgen}).

\section{Confining Lorentzian well via SUSY transformations}
We will show that the Lorentzian well model is shape invariant and that it can be mapped to the massless free particle system using supersymmetric transformations. This mapping is only possible when the coupling parameter of the Lorentzian well takes integer values, the same values that allow the confinement of zero-modes. Despite the fact that the Lorentzian well has been studied before \cite{Szafran,Mrenca,Downing, Downing2}, up to the authors knowledge, its shape invariant property has not been revealed yet. We also show how the zero modes appear due to the mapping between the two systems.

In order to proceed, let us multiply the equation in (\ref{DiracLorentz}) by $\Ti\sigma_2$. As a
result, we obtain a manifestly non-hermitian Hamiltonian operator $h_\ell$,
\begin{equation}
\label{geH}
h_\ell=\Ti\sigma_2 H_\ell\,,\qquad
h_\ell\psi=\left(\partial_y-\Ti \sigma_3\partial_x-\Ti\sigma_2\frac{2\ell}{1+x^2+y^2}\right)
\psi=0\,.
\end{equation}
Comparing $h_{\ell}$ with $h$ in (\ref{hgen}), we can identify $\gamma_1\equiv-\Ti \sigma_3$ and $\gamma_2=\sigma_0$. The new operator  $h_{\ell}$ preserves the symmetry represented by $\mathcal{T}\sigma_2$ while the solutions of (\ref{geH}) are the zero modes of the Lorentzian well.  It is worth noting that the Hamiltonian of the form $h_\ell$, belongs to the Davey-Stewartson II class of integrable systems \cite{Matveev}.  

As we discussed above, the Darboux transform is fixed by the matrix $U$, whose columns are eigenvectors of $h_\ell$. Here, since we are interested in designing a specific transformation whose properties will soon become clear, we fix $U$ in the following way  
\begin{equation}\label{U}
U_{\ell}=(\omega_{\ell},\mathcal{T}\sigma_2 \omega_{\ell}),\quad h_{\ell}U_{\ell}=0\,,
\end{equation}
where
\begin{align}\label{omega}
\omega_{\ell}=&\left(
\begin{array}{c}
-(x+\Ti y)^{-\ell-1 } \left(x^2+y^2+1\right)^{\ell} \\
\Ti (x+\Ti y)^{-\ell} \left(x^2+y^2+1\right)^{\ell} \\
\end{array}
\right)\,,\qquad
h_{\ell}\omega_{\ell}=0\,.
\end{align}
Once we have the matrix $U$ fixed, we can construct both the intertwining operator $L_{\ell}$ and the operator $\widetilde{h}_{\ell}$ following (\ref{Ltildeh}),
 \begin{align}
 \label{htildel}
 &\widetilde{h}_{\ell}=h_{\ell}-\Ti\begin{pmatrix}  \frac{x-\Ti y}{x+\Ti y} &0\\0&
 \frac{x+\Ti y}{x-\Ti y}\end{pmatrix}\sigma_2\frac{2}{1+x^2+y^2}\,,
  \\
 &L_{\ell}=\partial_x-\frac{1}{1+x^2+y^2}\left(
\begin{array}{cc}
 \frac{x^2+y^2-(\ell+1) (x^2+y^2-1)}{(x+\Ti y) } & \frac{\Ti x+2 (\ell+1) y-y}{(x+\Ti y)} \\
 \frac{\Ti x-2 (\ell+1) y+y}{(x-i y) } & \frac{x^2+y^2-(\ell+1) (x^2+y^2-1)}{(x-\Ti y) } \\
\end{array}
\right)\,, \\
& L_{\ell} h_{\ell}=\widetilde{h}_{\ell}L_{\ell}\,.\label{intetL}
 \end{align}
The new potential appearing in $\widetilde{h}_{\ell}$ has the shape of the Lorentzian well up to the multiplicative diagonal matrix.  We can transform out this matrix by a similarity transformation in terms of the matrix $S$ as follows, 
\begin{equation}
S \widetilde{h}_{\ell}S^{-1}=h_{\ell}-\Ti\sigma_2\frac{2}{1+x^2+y^2}=h_{\ell+1}\,,\qquad
\qquad
\widetilde{h}_{\ell}=S^{-1}h_{\ell+1} S\,,\qquad
S=\begin{pmatrix} x+\Ti y &0\\0& x-\Ti y\end{pmatrix}
\,.\label{S}
\end{equation}
When substituting $\widetilde{h}_{\ell}=S^{-1}h_{\ell+1} S$ 
into (\ref{intetL}), we get a new intertwining 
relation,
\begin{equation}
\label{intethl}
\mathcal{L}_{\ell}h_{\ell}=h_{\ell+1}\mathcal{L}_{\ell}\,.
\end{equation} 
Here, the operator $\mathcal{L}_{\ell}$ is a non-singular modified intertwining operator that satisfies $[\mathcal{T}\sigma_2,\mathcal{L}]=0$ and can be written explicitly,
\begin{equation}\label{calL}
\mathcal{L}_{\ell}=SL_{\ell}=
\left(
\begin{array}{cc}
(x+\Ti y)\partial_x+ 
 \ell-\frac{2\ell+1}{1+x^2+y^2} & -\frac{\Ti x+ (1+2\ell) y}{1+x^2+y^2} \\
 -\frac{\Ti x-(1+2\ell) y}{1+x^2+y^2} & (x-\Ti y)\partial_x +\ell-\frac{2\ell+1}{1+x^2+y^2} \\
\end{array}
\right)\,.
\end{equation} 
The intertwining relation (\ref{intethl}) indicates that the supersymmetric transformation (\ref{Ltildeh}) of the (non-hermitian) Hamiltonian $h_{\ell}$ produces exactly the same operator $h_{\ell+1}$ with the coupling constant shifted by one. Therefore, \textit{the operator $h_{\ell}$ is shape invariant} under the supersymmetric transformations generated by $\mathcal{L}_{\ell}$.

Up to now, we have derived the intertwining relations and revealed the shape invariance nature for the Hamiltonian $h_\ell$ in its non-hermitian form. Let us show what are the corresponding implications for the physically relevant Dirac Hamiltonian $H_\ell$. To derive the analogous intertwining relation for $H_\ell$, 
we can multiply (\ref{intethl}) from the left by $\sigma_2$, 
 \begin{equation}\label{LorHint}
\sigma_2\mathcal{L}_{\ell}\sigma_2 H_{\ell}=H_{\ell+1}\mathcal{L}_{\ell} \, .
\end{equation}
In contrast to (\ref{intethl}), this is an asymmetric intertwining relation for $H_{\ell}$ and $H_{\ell+1}$, i.e. the intertwining operator has a different form on the left side and on the right side. Such asymmetric intertwining relations have already appeared in the analysis of zero modes in \cite{cacj,Ioffe}. 

The Hamiltonian $H_{\ell}$ has $2\ell-2$ zero modes $\psi_{m,\ell}$ and   $\mathcal{T}\sigma_2\psi_{m,\ell}$ for $m=1,2, \dots \ell-1$. The intertwining relation can be used to map them to those of $H_{\ell+1}$,
\begin{align}
\label{mapp123}
H_{\ell}\psi_{m,\ell}&=0\qquad \Rightarrow \qquad H_{\ell+1}\mathcal{L}_{\ell}\psi_{m,\ell}=0\,, \qquad m=1,2,\dots \ell-1\, ,
\end{align}
while the states given by $\mathcal{T}\sigma_2\psi_{m,\ell}$ can be mapped in the same manner. In this way, we obtain the $2(\ell-1)$ zero modes for the Hamiltonian $H_{\ell+1}$. However, since the degeneracy of its zero energy is $2\ell$, there are two more missing zero modes to be found. To do this, we conjugate (\ref{intethl}) and multiply the relation by $\Ti\,\sigma_2$ in order to convert $h_\ell$ into $H_\ell$, 
\begin{equation}\label{shapeinvariance}
\mathcal{L}_{\ell}^{\dagger}H_{\ell+1}= H_{\ell} \sigma_2\mathcal{L}_{\ell}^\dagger\sigma_2\, .
\end{equation}
The two states annihilated by $\sigma_2\mathcal{L}_{\ell}^\dagger\sigma_2$ will be zero energy eigenstates of $H_\ell$. If they are square integrable, they will represent the missing bound states. 
The intertwining 
operator $\sigma_2\mathcal{L}_{\ell}^\dagger\sigma_2$ can be written in the following form 
\begin{equation}
\sigma_2\mathcal{L}_{\ell}^\dagger\sigma_2=-\sigma_2(U^{-1}_\ell)^\dagger\partial_xU_\ell^{\dagger} S^\dagger\sigma_2, 
\end{equation}
where we used (\ref{calL}). It annihilates the matrix $\widetilde{U}_{\ell+1}$ that is defined as follows,
\begin{equation}
\label{missingL}
\widetilde{U}_{\ell+1}=\sigma_2 (S^\dagger)^{-1}(U_{\ell}^\dagger)^{-1}=(\psi_{\ell,\ell+1},\mathcal{T}\sigma_2\psi_{\ell,\ell+1})\,.
\end{equation}
As the columns $\psi_{\ell,\ell+1}$  and $\mathcal{T}\sigma_2\psi_{\ell,\ell+1}$ are square integrable, see (\ref{spinorm}) they correspond to the two additional zero modes of $H_{\ell+1}$. 

Let us summarize our results so far. There is a mapping between $H_{\ell}$ and $H_{\ell+1}$ that can be used to obtain $2\ell-1$ zero modes of $H_{\ell+1}$ from those of $H_{\ell}$. The additional two zero modes are also fixed by the mapping, see (\ref{missingL}). This also explains the fact that the zero energy has even degeneracy, as the zero modes appear in pairs when the strength of the confining Lorentzian well coupling is increased from $\ell$ to $\ell+1$. In particular, the Hamiltonian $H_1$ has no normalizable zero modes that could be transformed into those of $H_2$. Nevertheless, the formula (\ref{missingL}) makes it possible to recover the pair of the zero modes, namely  $\psi_{1,2}$ and $\mathcal{T}\sigma_2\psi_{1,2}$. If we now perform the Darboux transformation on $H_2$, we get the Hamiltonian $H_3$. It inherits two zero modes from $H_2$ and the other two are generated by (\ref{missingL}), covering the expected fourfold degeneracy of the zero energy. 

The action of the intertwining operators on the zero modes can be obtained in a straightforward way using the Gauss contiguous relations for hypergeometric functions \cite{SpecialFunctions}. They provide the following relations
\begin{equation}
\label{rissinglow}
 \mathcal{L}_{\ell} \psi_{m,\ell}=(\ell+m+1)\psi_{m,\ell+1}\,,\qquad
 \widetilde{\mathcal{L}}_{\ell}\psi_{m,\ell}=(\ell-m-1)\psi_{m,\ell-1} \,,
\end{equation} 
where we denoted $\widetilde{\mathcal{L}}_{\ell}=\sigma_2\mathcal{L}_{\ell-1}^\dagger\sigma_2$
the intertwining operator that maps $H_\ell$ into $H_{\ell-1}$. These equations reveal that the Darboux transformations acts as a ladder operators with respect to the index $\ell$. The fact that the index $m$ is preserved by $\mathcal{L}_{\ell}$ is due to the following relations
\begin{equation}
[J,\mathcal{L}_{\ell}]=-
\mathcal{S}\sigma_2H_{\ell}\,,\qquad
[J,\widetilde{\mathcal{L}}_{\ell}]=
\mathcal{S}\sigma_2H_{\ell}\,.
\end{equation}
Since we are interested in the action of $\mathcal{L}_{\ell}$ in the subspace of zero modes, the right-hand sides of the equalities cancel out, and we find that $\mathcal{L}_{\ell}$ and 
$J$ effectively commute in this subspace.

The relations (\ref{rissinglow}) are valid for any integer value of $m$, including for the non-physical states.  In particular, the non-physical spinor $\psi_{\ell,\ell}$ of $H_{\ell}$ transforms into $\psi_{\ell,\ell+1}$. This state coincides with the first missing state in (\ref{missingL}). Thus, the non-physical state is regularized into a physical one by the transformation.

Let us now discuss the relation of $H_{\ell}$ to the free particle system. If we define a chain of Darboux transformations given by the operator, 
\begin{equation}
\label{higherOrderL}
\mathbb{L}_{\ell}=\mathcal{L}_\ell\mathcal{L}_{\ell-1}\dots\mathcal{L}_0 \, ,
\end{equation}
the Hamiltonian $H_\ell$ are intertwined with the free particle Hamiltonian $H_0$ by the following relations
\begin{equation}\label{freeparticle}
\mathbb{L}_{\ell} H_0=H_{\ell+1}\sigma_2\mathbb{L}_{\ell}\sigma_2 \, ,\quad  H_0\mathbb{L}_{\ell}^\dagger=\sigma_2\mathbb{L}_{\ell}^\dagger\sigma_2H_{\ell+1} \, .
\end{equation}
The kernel of the operator $\mathbb{L}_{\ell}$ consists of $2\ell$ states.  To find them, let us consider the non-physical solutions of $H_{\ell}$,
\begin{align}
\phi_{m,\ell}=\frac{(1+x^2+y^2)^\ell}{(x+\Ti y)^{m}  }
\left(
\begin{array}{c}
 -\frac{m}{x+\Ti y}  \, _2F_1\left(\ell,\ell-m,-m;-x^2-y^2\right) \\
 \Ti \ell \, _2F_1\left(\ell+1,\ell-m,1-m;-x^2-y^2\right) \\
\end{array}
\right)
\,,\qquad m=\ell,\ell+1,\ldots
\end{align}
where $\phi_{\ell,\ell}=\ell \omega_\ell$ is just the initial seed state, see (\ref{omega}).
Using the $\ell$ ladder operators, we find that 
\begin{equation}
\phi_{\ell,\ell}\propto\mathcal{L}_{\ell-1}\ldots\mathcal{L}_{0}\phi_{\ell,0}\,,\qquad
\mathcal{L}_{\ell}\phi_{\ell,\ell}=0\,.
\end{equation}
The state $\phi_{\ell-1,0}$ is annihilated by $\mathcal{L}_{\ell-1}\ldots\mathcal{L}_{0}\phi_{\ell,0}$, while the state $\phi_{\ell-1,0}$ vanishes after the action of $\mathcal{L}_{\ell-2}\ldots\mathcal{L}_{0}\phi_{\ell,0}$, and so on. In this way, the kernel of $\mathbb{L}_{\ell}$ can be written as

\begin{equation}
\text{Ker}(\mathbb{L}_{\ell})=\text{span}\{\phi_{0,0},\mathcal{T}\sigma_2\phi_{0,0},
 \phi_{1,0},\mathcal{T}\sigma_2 \phi_{1,0}\ldots\,\mathcal{T}\sigma_2  \phi_{\ell,0}\}\,,\qquad
\phi_{\ell,0}=\ell 
\left(
\begin{array}{c}
-(x+\Ti y)^{-\ell-1} \\
\Ti (-1)^\ell (x-\Ti y)^\ell \\
\end{array}
\right)\,. 
\end{equation}
Therefore, equation (\ref{freeparticle}) establishes the relation between the free particle zero energy solutions and those of the Lorentzian well. A visual scheme of this procedure is given in  Table \ref{tabstates}.
\begin{table}[H]
\begin{center}
\begin{tabular}{| c  c  c  c c c c|}
\hline 
$H_0\rightarrow$ & $H_1\rightarrow$ & $H_2\rightarrow$ & $H_3\rightarrow$& $H_4\rightarrow$ & \ldots & $H_\ell$ \\ \hline
\textcolor{red}{$\psi_{0,0}$} & \textcolor{red}{$\psi_{0,1}$}  & \textcolor{red}{$\psi_{0,2}$} & \textcolor{red}{$\psi_{0,3}$} & \textcolor{red}{$\psi_{0,4}$} & \ldots & \textcolor{red}{$\psi_{0,\ell}$}  \\
\textcolor{red}{$\psi_{1,0}$} & \textcolor{red}{$\psi_{1,1}$}  & \textcolor{blue}{$\psi_{1,2}$} & \textcolor{blue}{$\psi_{1,3}$} & \textcolor{blue}{$\psi_{1,4}$}& \ldots & \textcolor{blue}{$\psi_{1,\ell}$} \\
\textcolor{red}{$\psi_{2,0}$} & \textcolor{red}{$\psi_{2,1}$}  & \textcolor{red}{$\psi_{2,2}$} & \textcolor{blue}{$\psi_{2,3}$}& \textcolor{blue}{$\psi_{2,4}$}& \ldots & \textcolor{blue}{$\psi_{2,\ell}$}
\\
\textcolor{red}{$\psi_{3,0}$} & \textcolor{red}{$\psi_{3,1}$}  & \textcolor{red}{$\psi_{3,2}$} & \textcolor{red}{$\psi_{3,3}$}& \textcolor{blue}{$\psi_{3,4}$}  & \ldots &  \textcolor{blue}{$\psi_{3,\ell}$}
\\
. & .  & . & .&.  & \ldots & .
\\
. & .  & . & .&.  & \ldots & .
\\
. & .  & . & .&.  & \ldots & .
\\
. & .  & . & .&.  & \ldots & .
\\
\textcolor{red}{$\psi_{\ell-1,0}$} & \textcolor{red}{$\psi_{\ell-1,1}$}  & \textcolor{red}{$\psi_{\ell-1,2}$} & \textcolor{red}{$\psi_{\ell-1,3}$}& \textcolor{red}{$\psi_{\ell-1,4}$}  & \ldots &  \textcolor{blue}{$\psi_{\ell-1,\ell}$}
\\ \hline
\end{tabular}
\caption{\small{Schematic representation of the Darboux transformation (\ref{calL}) applied to each Hamiltonian. The arrows represent the application of the transformation, while the blue and red colors represent the physical (normalizable) and non-physical states of the form (\ref{spinorm}), 
respectively. Spinors of the form $\mathcal{T}\sigma_2\psi_{m,\ell}$ follow the same transformation rules, and for this reason, the total degeneracy is $2(\ell-1)$. Starting from the free case $H_0$ with no physical states of zero energy, the transformations gradually produce the Lorentzian well models for different values of  $\ell$ introducing at the same time new bound states.}}
\label{tabstates}
\end{center}
\end{table} 
  These results are similar to the situation of the intertwining relation between the non-relativistic free particle Hamiltonian and the energy operators of reflectionless models, see \cite{Correa, Plyushchay:2020yuw}. Nevertheless, the fact that the relation (\ref{freeparticle}) is asymmetric restricts its use to the subspace of zero energy solutions. 

The intertwining relations (\ref{intgen}) and (\ref{intgen2}) for the non-hermitian $h_{\ell}$ and $h_{\ell+1}$ can be rewritten in terms of the commuting extended matrix operators,
\begin{equation}\label{susyops}
\mathcal{H}_{\ell}=\begin{pmatrix}h_{\ell}&0\\0&h_{\ell+1}\end{pmatrix},\qquad \mathcal{Q}_{\ell}=\begin{pmatrix}0&\sigma_2\mathcal{L}_{\ell}^\dagger\sigma_2\\\mathcal{L}_{\ell}&0\end{pmatrix}\, , \qquad [\mathcal{H}_\ell,\mathcal{Q}_\ell]=0 \, .
\end{equation}
The commutation relations follow from (\ref{shapeinvariance}) and the conjugate relation with substitution $h_{\ell}^\dagger=-\sigma_2h_{\ell}\sigma_2$. It is remarkable that the diagonal operator $\mathcal{Q}_\ell^2$ commutes with the extended Hamiltonian $\mathcal{H}_\ell$ by means of the relations
\begin{equation}
[h_{\ell},\sigma_2\mathcal{L}_{\ell}^\dagger\sigma_2\mathcal{L}]=0,\quad [h_{\ell+1},\mathcal{L}_{\ell}\sigma_2\mathcal{L}_{\ell}^\dagger \sigma_2]=0 \, .
\end{equation}
The construction of the extended objects (\ref{susyops}) is inspired by Witten's model for the spontaneous breakdown of supersymmetry \cite{Cooper}, where the diagonal operator $\mathcal{H}_{\ell}$ represents the supersymmetric Hamiltonian while the anti-diagonal operator represents the supercharge $\mathcal{Q}_{\ell}$. In contrast to the toy model of Witten, the square of the supercharge operator does not coincide with the supersymmetric Hamiltonian, $\mathcal{Q}_{\ell}^2 \neq \mathcal{H}_{\ell}$. In our scenario, where the operators are also two-dimensional, the operator $\sigma_2\mathcal{L}_{\ell}^\dagger \sigma_2\mathcal{L}_{\ell}$ cannot be proportional to a polynomial in $h_{\ell}$, since there is no kinetic term of the form $\partial_y$. Despite the fact that this discussion holds for the non-hermitian system, we believe that it is an indication that hidden integrals for the hermitian system should somehow exist, and it would be interesting to find a way to reveal them.  It is worth  mentioning  in this context a series of papers where peculiar properties of physical systems have been related to the existence of pure quantum hidden symmetry operators constructed under a similar approach \cite{InzPly1,InzPly2}.

In the current two dimensional case the zero energy solutions of the free fermion case are given by spinors whose upper and lower component correspond to arbitrary functions of the form $f(x+\Ti y)$ and $g(x-\Ti y)$ respectively. This is why we can construct new models and solutions given in terms of the polar variables $r e^{\pm \Ti  \phi}=x\pm\Ti y$.

The shape invariance property means that the spinor
\begin{equation}
\label{generalSol1}
\Psi_{\ell+1}=\mathbb{L}_\ell \Psi_{\text{free}}\,,\qquad
\Psi_{\text{free}}=\left(\begin{array}{cc}
f(x+\Ti y)
\\
g(x-\Ti y)
\end{array}\right)
\end{equation}
is a zero energy solution of the Lorentzian well with parameter $\ell+1$. However, whether the resulting solutions are physical or not depends on the nature of the properties of the resulting vector $\mathbb{L}_\ell \Psi_{\text{free}}$ and thus on the choice of the functions $f(x+\Ti y)$ and $g(x-\Ti y)$. The freedom in the choice of the two functions opens the window for the search for a more general class of models with symmetries beyond radial, suitable for parabolic or spheroidal coordinates, for example.

\section{Conclusion}

In this article, we provided an alternative explanation for existence of zero modes in Lorenzian square well for integer values of the coupling parameter. We showed that the configurations that host the bound states are related by the supersymmetric transformation that intertwines the Hamiltonians with the coupling constants $2\ell$ and $2(\ell+1)$, revealing  shape invariance of these systems. In particular,  higher-order supersymmetric transformation (\ref{higherOrderL}) connects the confining Lorenzian well model with the free particle case $\ell=0$. This mapping explains the presence of zero energy bound states in the model and the $2(\ell-1)$-fold degeneracy of the zero energy. In addition, the general formula (\ref{generalSol1}) allows the straightforward construction of arbitrary zero modes from the states of free particles.

In this article, Darboux transformation was used in the way that deviates from its usual application. It was applied on the non-hermitian operator $h_\ell$ that was obtained from the Hamiltonian with Lorentzian well by a simple matrix manipulation, see (\ref{geH}). Additionally,  a similarity transformation (\ref{calL}) was incorporated into the definition of the intertwining operator $\mathcal{L}_\ell$, see (\ref{calL}). This way, the intertwining relations (\ref{intethl}) that recover the shape invariance of the operators  could be obtained. When recovering hermiticity of the Hamiltonians, the intertwining relation turned to be asymmetric and provided a mapping between the solutions with zero energies, see (\ref{LorHint}) and (\ref{shapeinvariance}). 

The innovative use of the supersymmetric transformation opens natural question whether it could be applied in similar manner for construction of other models. It would be interesting to consider systems in the related type of fields e.g. combination of Lorentzian wells. It is worth mentioning in this context that exact solutions for Dirac equation where Lorentzian well formed magnetic vector potential was analyzed in \cite{Nam1,Nam2}. One could seek for more models connected to the free particle by this modified transformation. Additionally, one can look for other examples of shape invariant cases. Even if the method works only for zero energy states, it could be adaptable and generalized for fixed energy levels with the redefinition of the potential by $V\rightarrow V-E$. Something similar was done in \cite{cacj}. In this vain, new potentials could be constructed with interesting observable properties, such as an anomalous hall effect\cite{susyhall} or a a non-trivial conductance \cite{waveguide1,waveguide2,waveguide3}. Of course, this is a line of research to explore.

It would be also interesting to consider systems with more particles. As much as the genuine many-body system with electron-electron interaction \cite{Wehling} are rather beyond the scope of our methods, we find it feasible to focus on the two-body systems whose effective Hamiltonian is written in terms of matrix formalism, see \cite{DowningBielectron} where formation of the bi-electron vortices was discussed. The non-hermitian operator $h_{\ell}$ appears as one of the linear auxiliary equations related to the nonlinear integrable Davey-Stewartson II equation \cite{Matveev}. In fact, the Lorentzian potential well can be understood in this context as a solitonic solution of the corresponding hierarchy. It is interesting to explore this relation in more detail, as well as the role of solitonic solutions from the point of view of fermionic models.  The exploration of these lines of research will be followed in the forthcoming publications.

\section*{Acknowledgement}
F.C. and L.I. were supported by Fondecyt Grants No.
1211356 and No. 3220327, respectively. V.J. acknowledges
the assistance provided by the Advanced Multiscale Materials
for Key Enabling Technologies project, supported by the Ministry of Education, Youth, and Sports of the Czech Republic.
Project No. CZ.$02.01.01/00/22\_008/0004558$, Co-funded by
the European Union. F.C. and L.I. are grateful for the hospitality of the Institute of Nuclear Physics of the Czech Academy
of Sciences and the Universidad Austral de Chile. All authors
also appreciate the hospitality of Cinvestav in Mexico.
 
\section{Trapping of nonrelativistic particles by Lorentzian well}
Let us suppose that we are dealing with a non-relativistic particle in the presence of a Lorentzian well described by the Hamiltonian
\begin{equation}
H_{\text{Sch}}=-\partial_x^2-\partial_y^2-\frac{V_0}{1+\frac{x^2+y^2}{d^2}} \, .
\end{equation}
The operator is hermitian and its spectrum is bounded from below. The continuous spectrum extends over the positive real half-line.  We can use the variational principle for the qualitative analysis of the ground state energy $E_0$. There holds $E_0\leq\frac{\langle \psi,H_{\text{Sch}}\psi\rangle}{\langle \psi,\psi\rangle}$, where $\psi$ is from the domain of $H_{\text{Sch}}$. 
When $V_0>0$, the attractive Lorentzian well can host bound states of negative energy. It can be proved by showing that the ground state energy must be negative.  Indeed, we can take a test function in the form of another Lorentzian function $\psi_{test}=\frac{\tilde{V}}{1+r^2/\widetilde{d}^2}$ and calculate the following integral,
\begin{equation}\label{var}
\langle \psi_{test},H_{\text{Sch}}\psi_{test}\rangle=\frac{2\pi\tilde{V}^2}{3}-\frac{d^2\tilde{d}^2\pi V_0\tilde{V}^2\left(d^2-\tilde{d}^2\left(1+2\ln d-2\ln \tilde{d}\right)\right)}{(d^2-\tilde{d}^2)^2}.
\end{equation} 
If $\langle \psi_{test},H_{\text{Sch}}\psi_{test}\rangle<0$, $E_0$ is below the threshold of the continuum spectrum. We can satisfy this inequality for sufficiently large values of $V_0$. Inserting the right-hand side of  (\ref{var}) into the latter inequality, we get the following result
\begin{equation}
V_0> \frac{2(d^2-\tilde{d}^2)^2}{3d^2\tilde{d}^2\left(d^2-\tilde{d}^2
\left(1+2\ln \frac{d}{\tilde{d}}\right)\right)} \, .
\end{equation}
	This means that for any $V_0$ that satisfies the inequality above,  there is at least one bound state with negative energy. This is qualitatively different from the case of Dirac Hamiltonian where only zero-energy bound states can exist and are hosted by the Lorenzian well for certain - discrete values of $V_0$. Therefore, the trapping of non-relativistic particles is more effective compared to Dirac fermions.


\vskip0.5cm

\begin{thebibliography}{99}


\bibitem{Tudorovski} T. Y. Tudorovskiy, A. V. Chaplik, 
\emph{``Spatially inhomogeneous states of charge carriers in graphene,"} 
\href{https://link.springer.com/article/10.1134/S002136400623010X}{JETP Lett. {\bf84}, 619  (2007).}


\bibitem{Downing3} C. A. Downing, M. E. Portnoi, \emph{``One-dimensional Coulomb problem in Dirac materials,"}
\href{https://journals.aps.org/pra/abstract/10.1103/PhysRevA.90.052116}{Phys. Rev. A {\bf90}, 052116 (2014)}
{\href{https://arxiv.org/abs/1411.5983}{\textcolor{magenta}{\tt[arXiv:1411.5983 [cond-mat.mes-hall]]}}}.


\bibitem{Hartmann}R. R. Hartmann, N. J. Robinson, and M. E. Portnoi, 
\emph{``Smooth electron waveguides in graphene,"}
\href{https://journals.aps.org/prb/abstract/10.1103/PhysRevB.81.245431}{Phys. Rev. B {\bf81}, 245431 (2010)}
{\href{https://arxiv.org/abs/0908.0561v3}{\textcolor{magenta}{\tt[arXiv:0908.0561 [cond-mat.mes-hall]]}}}.


\bibitem{Ho2}C.-L. Ho, P. Roy, 
\emph{``mKdV equation approach to zero energy states of graphene,"}
\href{https://iopscience.iop.org/article/10.1209/0295-5075/112/47004}{ EPL {\bf112}, 47004  (2015)}
{\href{https://arxiv.org/abs/1507.02649}{\textcolor{magenta}{\tt[arXiv:1507.02649 [quant-ph]]}}}.


\bibitem{Matulis}A. Matulis and F. M. Peeters, 
\emph{``Quasibound states of quantum dots in single and bilayer graphene,"}
\href{https://journals.aps.org/prb/abstract/10.1103/PhysRevB.77.115423}{Phys. Rev. B {\bf77}, 115423 (2008)}.
{\href{https://arxiv.org/abs/0711.4446v1}{\textcolor{magenta}{\tt[arXiv:0711.4446 [cond-mat.mes-hall]]}}}.


\bibitem{Downing} C. A. Downing, D. A. Stone, M. E. Portnoi, 
\emph{``Zero-energy states in graphene quantum dots and rings,"}
 \href{https://journals.aps.org/prb/abstract/10.1103/PhysRevB.84.155437}{Phys. Rev. B {\bf84}, 155437 (2011)}
 {\href{https://arxiv.org/abs/1105.0891}{\textcolor{magenta}{\tt[arXiv:1105.0891 [cond-mat.mes-hall]]}}}. 
 

\bibitem{Ioffe0}M. V. Ioffe, D. N. Nishnianidze,\emph{``Zero energy states for a class of two-dimensional potentials in graphene,"} 
\href{https://www.worldscientific.com/doi/abs/10.1142/S0217984918503293}{Mod. Phys. Lett. B {\bf32}, 1850329 (2018)}
{\href{https://arxiv.org/abs/1811.01637}{\textcolor{magenta}{\tt[arXiv:1811.01637 [cond-mat.mes-hall]]}}}.


\bibitem{Downing2} C. A. Downing and M. E. Portnoi, 
\emph{``Zero-Energy Vortices in Dirac Materials,"}
\href{https://onlinelibrary.wiley.com/doi/10.1002/pssb.201800584}{Phys. Status Solidi B {\bf 256}, 1800584 (2019)}
 {\href{https://arxiv.org/abs/1903.09005}{\textcolor{magenta}{\tt[arXiv:1903.09005 [cond-mat.mes-hall]]}}}.


\bibitem{Szafran}B. Szafran, \emph{``Scanning gate microscopy simulations for quantum rings: Effective potential of the tip and conductance maps,"} 
\href{https://journals.aps.org/prb/abstract/10.1103/PhysRevB.84.075336}{Phys. Rev B {\bf 84}, 075336 (2011)}. 

	
\bibitem{Mrenca} A. Mre\'nca, K. Kolasi\'ski, B. Szafran, \emph{``Imaging localization of quasibound states in graphene antidots,"}
\href{https://journals.aps.org/prb/abstract/10.1103/PhysRevB.90.035314}{Phys. Rev. B {\bf90}, 035314 (2014)}
{\href{https://arxiv.org/abs/1407.0479}{\textcolor{magenta}{\tt[arXiv:1407.0479 [cond-mat.mes-hall]]}}}. 
	
	
\bibitem{Cooper}	F.~Cooper, A.~Khare and U.~Sukhatme,
{\it Supersymmetry in quantum mechanics,}
(World Scientific, Singapore, 2001).


\bibitem{Junker} G. Junker, {\it Supersymmetric Methods in Quantum and Statistical Physics,} (Springer, Berlin, 1996)


\bibitem{Matveev} V. B. Matveev, M. A. Sale, \textit{Darboux Transformations and Solitons}, (Springer-Verlag, Berlin, 1991).


\bibitem{Samsonov}L. M. Nieto, A. A. Pecheritsin, B. Samsonov,
\emph{``Intertwining technique for the one-dimensional stationary Dirac equation,"}
\href{https://www.sciencedirect.com/science/article/pii/S000349160300071X}{Annal Phys. {\bf305}, 151 (2003)}.
{\href{https://arxiv.org/abs/quant-ph/0307152}{\textcolor{magenta}{\tt[arXiv:quant-ph/0307152]}}}.

\bibitem{Samsonov2} A. A. Pecheritsyn, E. E. Pozdeeva, B. F. Samsonov, \emph{``Darboux Transformation of the Nonstationary Dirac Equation,"} 
\href{https://link.springer.com/article/10.1007/s11182-005-0134-x}{Russian Physics Journal \textbf{48}, 365-374 (2005)}.

\bibitem{Jakubsky1}
V. Jakubsk\'y, M. Plyushchay,
\emph{``Supersymmetric twisting of carbon nanotubes,"}
\href{https://journals.aps.org/prd/abstract/10.1103/PhysRevD.85.045035}{Phys. Rev. D {\bf85}, 045035 (2012)}
{\href{https://arxiv.org/abs/1111.3776v2}{\textcolor{magenta}{\tt[arXiv:1111.3776 [hep-th]]}}}.


\bibitem{Correa}F. Correa, V. Jakubsk\'y, 
\emph{``Twisted kinks, Dirac transparent systems and Darboux transformations,"} 
\href{https://journals.aps.org/prd/abstract/10.1103/PhysRevD.90.125003}{Phys. Rev. D {\bf 90}, 125003 (2014)}
{\href{https://arxiv.org/abs/1406.2997v2}{\textcolor{magenta}{\tt[arXiv:1406.2997 [hep-th]]}}}.


\bibitem{Correa2} F. Correa, V. Jakubsk\'y, 
\emph{``Confluent Crum-Darboux transformations in Dirac Hamiltonians with $PT$-symmetric Bragg gratings,"}
\href{https://journals.aps.org/pra/abstract/10.1103/PhysRevA.95.033807}{Phys. Rev. A {\bf95}, 033807 (2017)}
{\href{https://arxiv.org/abs/1612.06349v1}{\textcolor{magenta}{\tt[arXiv:1612.06349 [hep-th]]}}}.

 \bibitem{Ho}C.-L. Ho and P. Roy, 
\emph{``On zero energy states in graphene,"}
 \href{https://iopscience.iop.org/article/10.1209/0295-5075/108/20004/meta}{EPL {\bf108}, 20004 (2014)}.
 
 
 \bibitem{Schulze1}A. Schulze-Halberg, P. Roy, 
 \emph{``Construction of zero-energy states in graphene through the supersymmetry formalism,"}  
 \href{https://iopscience.iop.org/article/10.1088/1751-8121/aa8249/meta}{J. Phys. A: Math. Theor. {\bf50}, 365205 (2017)}.
 
 
 \bibitem{Ghosh}
 P. Ghosh and P. Roy, 
 \emph{``An analysis of the zero energy states in graphene,"}
\href{https://www.sciencedirect.com/science/article/pii/S0375960115010415}{Phys. Lett. A {\bf380}, 567-569 (2016)}.

\bibitem{cacj}
A.~Contreras-Astorga, F.~Correa and V.~Jakubsky,
\emph{``Super-Klein tunneling of Dirac fermions through electrostatic gratings in graphene,''}
\href{https://doi:10.1103/PhysRevB.102.115429}{Phys. Rev. B \textbf{102}, no.11, 115429 (2020)}
{\href{https://arxiv.org/abs/2006.08207}{\textcolor{magenta}{\tt[arXiv:2006.08207 [cond-mat.mes-hall]]}}}.


\bibitem{susyhall}
M.~Ezawa, \emph{``Supersymmetry and unconventional quantum Hall effect in graphene,''}
\href{https://doi.org/10.1016/j.physleta.2007.08.071}{Phys. Lett. A \textbf{372}, 924-929 (2008)}
{\href{https://arxiv.org/abs/cond-mat/0606084}{\textcolor{magenta}{\tt[arXiv:0606084 [cond-mat.mes-hall]]}}}.

\bibitem{waveguide1} 
R. A. Ng, A. Wild, M. E. Portnoi and  R. R. Hartmann,
\emph{``Mapping borophene onto graphene: Quasi-exact solutions for guiding potentials in tilted Dirac cones,"}
\href{https://doi.org/10.1038/s41598-022-11742-3}{Sci Rep \textbf{12}, 7688 (2022)}
\href{https://arxiv.org/abs/2111.10760}{\textcolor{magenta}{\tt [arXiv:2111.10760 [cond-mat.mes-hall]]}}.


\bibitem{waveguide2} 
R. R. Hartmann, M. E. Portnoi,
\emph{``Quasi-exact solution to the Dirac equation for the hyperbolic secant potential,"}
\href{https://journals.aps.org/pra/abstract/10.1103/PhysRevA.89.012101}{Phys. Rev. A \textbf{89}, 012101 (2014)}
\href{https://arxiv.org/abs/1305.4652}{\textcolor{magenta}{\tt [arXiv:1305.4652 [cond-mat.mes-hall]]}}.



\bibitem{waveguide3} 
R. R. Hartmann, M. E. Portnoi
\emph{``Two-dimensional Dirac particles in a P\"oschl-Teller waveguide,"}
\href{https://www.nature.com/articles/s41598-017-11411-w}{Sci. Rep. \textbf{7}, 11599 (2017)}
\href{https://arxiv.org/abs/1709.07147}{\textcolor{magenta}{\tt [ arXiv:1709.07147]}}.

 

\bibitem{Dunne:2013yra}
G.~V.~Dunne and M.~Thies,  \emph{``Transparent dirac potentials in one dimension: the time-dependent case,''}
\href{https://doi:10.1103/PhysRevA.88.062115}{Phys. Rev. A \textbf{88}, no.6, 062115 (2013)} 
{\href{https://arxiv.org/abs/1308.5801}{\textcolor{magenta}{\tt[arXiv:1308.5801 [hep-th]]}}}.

\bibitem{dt2}
G.~V.~Dunne and M.~Thies, 
\emph{``Time-Dependent Hartree-Fock Solution of Gross-Neveu models: Twisted Kink Constituents of Baryons and Breathers,''}
\href{https://doi:10.1103/PhysRevLett.111.121602}{Phys. Rev. Lett. \textbf{111}, no.12, 121602 (2013)}
{\href{https://arxiv.org/abs/1306.4007}{\textcolor{magenta}{\tt[arXiv:1306.4007 [hep-th]]}}}.

\bibitem{Dutt}
R. Dutt, A. Khare and  U.P. Sukhatme,
\emph{``Supersymmetry, shape invariance, and exactly solvable potentials,"}
\href{https://pubs.aip.org/aapt/ajp/article-abstract/56/2/163/1038855/Supersymmetry-shape-invariance-and-exactly?redirectedFrom=fulltext}{American Journal of Physics {\bf 56}, 163 (1988)}.

\bibitem{SpecialFunctions}
F. W. Olver, D. W. Lozier, R. F.  Boisvert and  
 C. W. Clark, \textit{NIST handbook of mathematical functions}  
 (Cambridge university press, 2010).

\bibitem{Hsu} 
Ch. W. Hsu, B. Zhen, A. D. Stone, J. D. Joannopoulos, and M. Solja\v ci\'c,
\emph{``Bound states in the continuum,"}
 \href{https://www.nature.com/articles/natrevmats201648}{Nat. Rev. Mater. \textbf{1}, 16048 (2016)}.

\bibitem{Gonzalez} J. W. Gonz\'alez, M. Pacheco, L. Rosales, and P. A. Orellana, 
\emph{``Bound states in the continuum in graphene quantum dot structures,"}
\href{https://iopscience.iop.org/article/10.1209/0295-5075/91/66001}{Europhys. Lett. \textbf{91}, 66001 (2010)}
\href{https://arxiv.org/abs/1009.2711}{\textcolor{magenta}{\tt[arXiv:1009.2711 [cond-mat.mes-hall]]}}.

\bibitem{Cortes} 
N. Cort\'es, L. Chico, M. Pacheco, L. Rosales, and P. A. Orellana,
 \emph{``Bound states in the continuum: Localization of Dirac-like fermions,"}
 \href{https://iopscience.iop.org/article/10.1209/0295-5075/108/46008/meta}{ Europhys. Lett. \textbf{108}, 46008 (2014)}
\href{https://arxiv.org/abs/1406.2627}{\textcolor{magenta}{\tt[arXiv:1406.2627 [cond-mat.mes-hall]]}}.

\bibitem{JLee}
J. Lee  et al, \emph{``Imaging electrostatically confined Dirac fermions in graphene quantum dots,"} 
\href{https://www.nature.com/articles/nphys3805}{Nature Physics \textbf{12.11} (2016): 1032-1036}.

\bibitem{Freitag}
 N. M. Freitag  et al. \emph{``Electrostatically confined monolayer graphene quantum dots with orbital and valley splittings,"} 
 \href{https://pubs.acs.org/doi/full/10.1021/acs.nanolett.6b02548}{Nano letters \textbf{16.9} (2016): 5798-5805}.


\bibitem{LiLin}
Si-Yu Li and Lin He, 
\emph{``Recent progresses of quantum confinement in graphene quantum dots,"}
\href{https://link.springer.com/article/10.1007/s11467-021-1125-2}{ Frontiers of Physics \textbf{17} (2022): 1-25}.


\bibitem{Yung}
K.C. Yung, W.M. Wu, M.P. Pierpoint, F.V. Kusmartsev
\emph{``Introduction to graphene electronics -- a new era of digital transistors and devices,"}
\href{https://www.tandfonline.com/doi/abs/10.1080/00107514.2013.833701}
{Contemporary Physics \textbf{54.5} (2013): 233-251}
\href{https://arxiv.org/abs/1309.0205v1}{\textcolor{magenta}{\tt[arXiv:1309.0205 [cond-mat.mtrl-sci]]}}.



\bibitem{Darboux}G. Darboux, \emph{``Sur une proposition relative auxequations lineaires," }
Compt. Rend. Acad. Sci. (Paris) {\bf94}, 1456-1459 (1882).





 \bibitem{Ioffe}M. V. Ioffe, D. N. Nishnianidze, E. V. Prokhvatilov,
\emph{``New solutions for graphene with scalar potentials by means of generalized intertwining," }
\href{https://epjplus.epj.org/articles/epjplus/abs/2019/09/13360_2019_Article_12798/13360_2019_Article_12798.html}{Eur. Phys. J. Plus {\bf 134}, 450 (2019)}.


\bibitem{Plyushchay:2020yuw}
M.~S.~Plyushchay,
\emph{``Exotic Nonlinear Supersymmetry and Integrable Systems,''}
\href{https://doi:10.1134/S1063779620040589}{Phys. Part. Nucl. \textbf{51}, no.4, 583-588 (2020)}
{\href{https://arxiv.org/abs/2001.04133}{\textcolor{magenta}{\tt[arXiv:2001.04133 [hep-th]]}}}.

 

 





\bibitem{InzPly1}
L.~Inzunza and M.~S.~Plyushchay,
\emph{``Hidden symmetries of rationally deformed superconformal mechanics,''}
\href{https://journals.aps.org/prd/abstract/10.1103/PhysRevD.99.025001}{Phys. Rev. D \textbf{99} (2019), 025001}
{\href{https://arxiv.org/abs/quant-ph/0307152}{\textcolor{magenta}{\tt[arXiv:1809.08527 [hep-th]]}}}.



\bibitem{InzPly2}
L.~Inzunza and M.~S.~Plyushchay,
\emph{``Klein four-group and Darboux duality in conformal mechanics,''}
\href{https://journals.aps.org/prd/abstract/10.1103/PhysRevD.99.125016}{Phys. Rev. D \textbf{99} (2019), 125016}
{\href{https://arxiv.org/abs/quant-ph/0307152}{\textcolor{magenta}{\tt [arXiv:1902.00538 [hep-th]]}}}.


\bibitem{Nam1}D-N. Le, V-H. Le, P. Roy,\emph{"Conditional electron confinement in graphene via smooth magnetic fields,"} 
\href{https://www.sciencedirect.com/science/article/abs/pii/S1386947717313292}{Physica E {\bf96}, 17 (2018)}
\href{https://arxiv.org/abs/1711.04509}{\textcolor{magenta}{\tt[arXiv:1711.04509 [cond-mat.mes-hall]]}}.


\bibitem{Nam2}D-N. Le, V-H. Le, P. Roy, \emph{"Generalized harmonic confinement of massless Dirac fermions in (2+1) dimensions,"} 
\href{https://www.sciencedirect.com/science/article/abs/pii/S1386947718304351}{
Physica E {\bf102}, 66 (2018)}
\href{https://arxiv.org/abs/1804.08268}{\textcolor{magenta}{\tt[arXiv:1804.08268 [cond-mat.mes-hall]]}}.


\bibitem{Wehling} T. O. Wehling  E. Sasioglu, C. Friedrich, A. I. Lichtenstein, M. I. Katsnelson and  S. Bl\"ugel, \emph{``Strength of Effective Coulomb Interactions in Graphene and Graphite,"} \href{https://journals.aps.org/prl/abstract/10.1103/PhysRevLett.106.236805}{Phys. Rev. Lett. \textbf{106}, 236805 (2011)}
\href{https://arxiv.org/abs/1101.4007}{\textcolor{magenta}{\tt[arXiv:1101.4007 [cond-mat.mes-hall]]}}.


\bibitem{DowningBielectron} C. A. Downing, M. E. Portnoi, 
\emph{``Bielectron vortices in two-dimensional Dirac semimetals,"} 
\href{https://www.nature.com/articles/s41467-017-00949-y}{Nat. Commun. \textbf{8}, 897 (2017)}
\href{https://arxiv.org/abs/1506.04425}{\textcolor{magenta}{\tt[arXiv:1506.04425 [cond-mat.mes-hall]]}}.




\end{thebibliography}
\end{document}